\documentstyle[aps,12pt]{revtex}

\begin{document}
\draft
\author{O. B. Zaslavskii}
\address{Department of Physics, Kharkov State University, Svobody Sq.4, Kharkov\\
310077, Ukraine\\
E-mail: aptm@kharkov.ua}
\title{Two- and Many-Dimensional Quasi-Exactly Solvable Models With An\\
Inhomogeneous Magnetic Field}
\maketitle

\begin{abstract}
{Let group generators having finite-dimensional representation be realized
as Hermitian linear differential operators without inhomogeneous terms as
takes place, for example, for the SO(n) group. Then corresponding group
Hamiltonians containing terms linear in generators (along with quadratic
ones) give rise to quasi-exactly solvable models with a magnetic field in a
curved space. In particular, in the two-dimensional case such models are
generated by quantum tops. In the three-dimensional one for the SO(4)
Hamiltonian with an isotropic quadratic part the manifold within which a
quantum particle moves has the geometry of the Einstein universe.}
\end{abstract}

\address{Department of Physics, Kharkov State University, 4 Svobody Square,\\
Kharkov 310077, Ukraine\\
E-mail: aptm@kharkov.ua}



Exact solutions with a magnetic field are extremely rare in quantum
mechanics. One can find only two examples of such a kind in textbooks: a
free electron or an harmonic oscillator, a magnetic field being homogeneous.
Below we show how that in fact there exists a variety of systems with exact
solutions of the Schr\"{o}dinger equation with an inhomogeneous magnetic
field among quasi-exactly solvable models (QESM). In so doing, not only a
potential and magnetic field are present but also the analog of a
gravitational one in that a corresponding particle is moving on a curved
surface \cite{zasl1}, \cite{zasl2}.

First let us consider the two-dimensional case following a general procedure 
\cite{shif}. We construct Hamiltonian which contains quadratic and linear
terms in generators of a Lie-algebra having a finite-dimensional
representation and choose the algebra of the SO(3) group, i.e. algebra of an
angular momentum operators: 
\begin{equation}
H=\alpha L_{x}^{2}+\beta L_{y}^{2}+\gamma
L_{z}^{2}+C_{1}L_{x}+C_{2}L_{x}+C_{3}L_{z}\text{,}  \label{ham}
\end{equation}
$L_{x}=i(\sin \phi \frac{\partial }{\partial \theta }+\cos \phi ctg\theta 
\frac{\partial }{\partial \phi })$, $L_{y}=i(-\cos \phi \frac{\partial }{%
\partial \theta }+\sin \phi ctg\theta \frac{\partial }{\partial \phi })$, $%
L_{z}=-i\frac{\partial }{\partial \phi }$, $\theta $ and $\phi $ are the
angles of the spherical coordinate system.

Substitute these expressions into the Schr\"{o}dinger equation 
\begin{equation}
H\Phi =E\Phi \text{.}  \label{Sch}
\end{equation}
It can be represented in the form 
\begin{equation}
-g^{\mu \nu }\frac{\partial ^{2}\Phi }{\partial X^{\mu }\partial X^{\nu }}%
+T^{\mu }\frac{\partial \Phi }{\partial X^{\mu }}=E\Phi \text{,}  \label{dif}
\end{equation}
$\mu $, $\nu =\theta $,$\phi $.

This can be rewritten as follows, 
\begin{equation}
-g^{\mu \nu }(\nabla _{\mu }-A_{\mu })(\nabla _{\nu }-A_{\nu })\Phi +U\Phi
=E\Phi  \label{cov}
\end{equation}
where $\nabla $ is the covariant derivative operator. Comparing (\ref{dif})
and (\ref{cov}) one can find the explicit expression for $A_{\mu }$. In
order for the equation (\ref{cov}) to take this form it is necessary that in
the expression for $A_{\mu }$, 
\begin{equation}
A_{\mu }=a_{\mu }+ib_{\mu }\text{,}  \label{a}
\end{equation}
its real part be a pure gradient: $a_{\mu }=\rho _{,\mu }$. Then, by
substitution $\Psi =\Phi e^{-\rho }$ eq. (\ref{cov}) is reduced to the form 
\begin{equation}
-g^{\mu \nu }(\nabla _{\mu }-ib_{\mu })(\nabla _{\nu }-ib_{\nu })\Psi +U\Psi
=E\Psi \text{.}  \label{eq}
\end{equation}
Imaginary terms in (\ref{eq}) can be attributed in a natural way to the
two-dimensional analog of a magnetic field (or its component orthogonal to
the surface if the model is thought of as embedded into three-dimensional
space). The value of this field, $B=\frac{b_{\theta ,\phi }-b_{\phi ,\theta }%
}{\sqrt{g}}$ ($g=\det g_{\mu \nu }$) determines the value of the
two-dimensional field invariant $B^{2}=\frac{1}{2}F_{\mu \nu }F^{\mu \nu }$, 
$F_{\mu \nu }=b_{\nu ,\mu }-b_{\mu ,\nu }$. Comparing (\ref{dif}) and (\ref
{cov}) one can find the potential $U=g_{\mu \nu }A^{\mu }A^{\nu
}-g^{-1/2}(g^{1/2}A^{\mu })_{,\mu }$ which is in general complex: $%
U=U_{1}+iU_{2}$, 
\begin{equation}
U_{2}=2a_{\mu }b^{\mu }-g^{-1/2}(g^{1/2}a^{\mu })_{,\mu }.  \label{u}
\end{equation}

If $b_{\mu }=0$ we return to the situation discussed in detail in \cite{shif}%
. In that case the condition of integrability 
\begin{equation}
a_{\mu }=\rho _{,\mu }  \label{i}
\end{equation}
is sufficient for the equation under consideration to take the
Schr\"{o}dinger form. However, if $b_{\mu }\neq 0$ the second condition $%
U_{2}=0$ is called, otherwise the Hamiltonian would become non-Hermitian. In
general, one could expect that the imaginary part of the potential $%
U_{2}\neq 0$ which is the obstacle for obtaining a physically reasonable
QESM with a magnetic field. A nontrivial point in the problem under
consideration is that for the S)(3) Hamiltonian the potential turns out to
be purely real as it follows after some direct but lengthy calculations. In
general the expression for the potential is rather cumbersome, so we list
only some simplest examples.

In the isotropic case $\alpha =\beta =\gamma $ one may choose the coordinate
system in such a way that $C_{1}=C_{2}=0$, $C_{3}=C$. Then $U=-\frac{%
C^{2}\sin ^{2}\theta }{4}$, $B=C\cos \theta $, $g_{\theta \theta }=\alpha
^{-1}$, $g_{\phi \phi }=\alpha ^{-1}\sin ^{2}\theta $. The manifold now is
nothing but a sphere. If $\alpha =\beta $, $\gamma =0$, $C_{2}=0$ the
Riemannian curvature $R=-4\alpha \cos ^{-2}\theta $, $B=C_{3}\cos
^{-2}\theta -\frac{C_{1}}{2}\cos \phi tg\theta $.

The manifold within which a quantum particle moves is compact if $\alpha $, $%
\beta $, $\gamma >0$ and noncompact if one of these constants is zero. It
turns out that the wave function is normalizable even in cases (like in the
example above) when curvature or potential can contain singularities.

Note that since the square of the angular momentum commutes with the
Hamiltonian (\ref{ham}) all the space of states can be divided into
subspaces with fixed values of angular momentum $l$. In turn, in any such a
subspace the Hamiltonian (\ref{ham}) in a matrix representation is
equivalent to a spin Hamiltonian with spin $l$ and generates a
one-dimensional QESM with a potential composed of elliptic functions\cite{ul}%
. This is rather interesting correspondence between the one-dimensional QESM
and two-dimensional ones defined on curved surfaces and with a magnetic
field. For small values of $l$ the expressions for energy levels and wave
functions can be found explicitly. For an arbitrary $l$ they enter into the
algebraic equation of a finite degree that is typically for QESM.

It is the generalization of the obtained results to the many-dimensional
case that we now turn to. Consider group Hamiltonian 
\begin{equation}
H=C_{ab}L^{a}L^{b}+C_{a}L^{a}  \label{group}
\end{equation}
with real coefficients, $C_{ab}=C_{ba}$. In fact, we use, as the starting
point, approach of \cite{mor} where it was shown that the choice $C_{a}=0$
leads to both hermiticity of (\ref{group}) with a certain measure in
Riemannian space and an absence of a magnetic field. Below we show that
taking $C_{a}\neq 0$ also preserves hermiticity and corresponds to the
appearance of a certain magnetic field of the Schr\"{o}dinger operator. Let
the generators have the form 
\begin{equation}
L^{a}=ih^{a\mu }\frac{\partial }{\partial x^{\mu }}  \label{gen}
\end{equation}
with real $h^{a\mu }$. Substituting (\ref{gen}) into the Schr\"{o}dinger
equation we obtain the second order differential equation which can be
rewritten in the form (\ref{cov}) with $g^{\mu \nu }=C_{ab}h^{a\mu }h^{b\nu
} $, $T^{\mu }=-C_{ab}h^{a\nu }h_{,\nu }^{b\mu }+h^{a\mu }C_{a}$, $A^{\mu }=%
\frac{1}{2}(P^{\mu }+T^{\mu })$, $P^{\mu }\equiv $ ($\sqrt{g}g^{\mu \nu }),/%
\sqrt{g}$, $U_{2}$ having the same form as in the two-dimensional case. The
crucial point is whether or not $U_{2}=0$. The real part of the above
expression for $A_{\mu }$ gives us 
\begin{equation}
C_{ab}h^{a\mu }(2h^{b\nu }a_{\nu }-h^{b\nu }\frac{\sqrt{g}_{,\nu }}{\sqrt{g}}%
-h_{,\nu }^{b\nu })=0\text{.}  \label{16}
\end{equation}

Now we invoke an additional assumption \cite{mor}:\ let operators $L_{a}$ be
Hermitian in some metric $g_{\mu \nu }^{(0)}$ in which the scalar product is
determined in a standard way: 
\begin{equation}
(\phi _{2}\text{,}\phi _{1})=\int d^{n-1}x\sqrt{g^{(0)}}\phi _{2}^{*}\phi
_{1}\text{.}  \label{prod}
\end{equation}
For example, for the $SO(n)$ group $g_{\mu \nu }^{(0)}$ is the metric of a $%
n-1$ dimensional hypersphere. Then the hermiticity condition 
\begin{equation}
(\phi _{2}L^{\mu }\phi _{1})=(L^{a}\phi _{2}\text{, }\phi _{1})  \label{herm}
\end{equation}
along with (\ref{gen}) and (\ref{prod}) entails 
\begin{equation}
h^{a\mu }\text{,}_{\mu }=-h^{a\mu }\frac{\partial }{\partial x^{\mu }}\ln 
\sqrt{g^{(0)}}  \label{h}
\end{equation}
It follows from (\ref{16}), (\ref{h}) that 
\begin{equation}
C_{ab}h^{a\mu }h^{b\nu }(2a_{\nu }-\frac{\sqrt{g}_{,\nu }}{\sqrt{g}}+\frac{%
\sqrt{g(0)}_{,\nu }}{\sqrt{g(0)}})=0\text{.}  \label{20}
\end{equation}
It is clear that irrespective of $C_{ab}$ there exists the solution $a_{\mu
}=\rho _{,\nu }$ with $\rho =\frac{1}{2}\ln \sqrt{g/g^{(0)}}$. After the
substitution $\Psi =\Phi e^{-\rho }$ the Schr\"{o}dinger equation takes the
form (\ref{eq}), $\Psi $ being normalized according to $(\Psi ,\Psi )=\int
d^{n-1}x\sqrt{g}\left| \Psi \right| ^{2}=\int d^{n-1}\sqrt{g^{(0)}}\left|
\Phi \right| ^{2}$. Thus, normalizability of $\Phi $ entails normalizability
of $\Psi $.

Until now our treatment has run almost along the same lines as in \cite{mor}
where it was assumed $C_{a}=0$. The key new moment which makes our problem
non-trivial is that for $C_{a}\neq 0$ the potential contains the imaginary
part $iU_{2}$ and one must elucidate whether or not $U_{2}=0$. Using
explicit formulae for the coefficients of the differential equations listed
above one can show that 
\begin{equation}
2a_{\mu }b^{\mu }=\frac{1}{2}h^{a\mu }C_{a}(\frac{\sqrt{g}_{,\mu }}{\sqrt{g}}%
-\frac{\sqrt{g^{(0)}}_{,\mu }}{\sqrt{g^{(0)}}})\text{,}  \label{ab}
\end{equation}
\begin{equation}
\frac{(b^{\mu }\sqrt{g})_{,\mu }}{\sqrt{g}}=b_{,\mu }^{\mu }+\frac{\sqrt{g}%
_{,\mu }}{\sqrt{g}}b^{\mu }=\frac{C_{a}}{2}(h_{,\mu }^{a\mu }+\frac{\sqrt{g}%
_{,\mu }}{\sqrt{g}}h^{a\mu })  \label{b}
\end{equation}
Making use of (\ref{h}), we see that expressions (\ref{ab}) and (\ref{b})
coincide completely, so according to (\ref{u}) $U_{2}=0!$

Thus, the general form of generators (\ref{gen}) along with the hermiticity
condition (\ref{prod}) and (\ref{herm}) entail the integrability condition (%
\ref{i}) and, simultaneously, ensure that the potential is real. In other
words, if generators (\ref{gen}) Hermitian in a space with the metric $%
g_{\mu \nu }^{(0)}$ , Hamiltonian of the Schrodinger equation is Hermitian
in a space with the metric $g_{\mu \nu }$.

The result obtained shows that there is essential difference between QESM\
based on SO(n) groups and, say, SU(n) ones. In the latter case even when a
magnetic field is absent it is rather difficult task to find coefficients $%
C_{ab}$ for which the integrability condition (\ref{i}) is satisfied \cite
{shif}. In the second case we obtain at once QESM with well defined
Hamiltonian for which this condition is satisfied automatically and,
moreover, the effective magnetic field $F_{\mu \nu }=b_{\nu ,\mu }-b_{\mu
,\nu }$ is present.

Consider briefly an example of QESM of such a kind in three-dimensional
space based on generators of SO(4) $L_{ik}=-i(x^{i}\partial /\partial
x^{k}-x^{k}\partial /\partial x^{i})$. For Hamiltonian $H=%
\sum_{i<k}L_{ik}^{2}+CL_{12}$ direst calculations show that the potential
and field tensor are equal to $U=-\frac{C^{2}}{2}\sin ^{2}\xi \sin
^{2}\theta $, $F_{\theta \phi }=-C\sin ^{2}\xi \sin \theta \cos \theta $, $%
F_{\xi \phi }=-C\sin ^{2}\xi \sin \xi \cos \xi $, $F_{\xi \theta }=0.$ The
metric reads $ds^{2}=d\theta ^{2}+\sin ^{2}\theta d\xi ^{2}+\sin ^{2}\theta
\sin ^{2}\xi d\phi ^{2}$ where $\theta $,$\xi $,$\phi $ are angles of the
hyperspherical coordinate systems. The obtained metric is nothing else than
that of the Einstein universe.

It is worth noting that the approach outlined above enables one to obtain
QESM with a magnetic field which are not reduced to exactly solvable ones.
Whereas for exactly solvable cases such a charged particle or an harmonic
oscillator in an homogeneous magnetic field finding exact solutions implies
separation of variables, such separation is not needed for solutions under
discussion. In so doing, we obtain many-parametric classes of solutions at
once, a magnetic field being inhomogeneous. The essential feature of QESM in
question is that a manifold on which a quantum particle moves is inevitable
curved. This can be of interest, for example, for applications in
relativistic cosmology.

\section*{Acknowledgments}

This work is supported by International Science Education Program, grant \#
QSU082068.





%
%

%
%


\section*{References}

\begin{references}
\bibitem{zasl1}  O. B. Zaslavskii, {\em Physics Letters} {\bf A 190}, 373
(1994).

\bibitem{zasl2}  O. B. Zaslavskii, {\em Journal of Physics} {\bf A 27}, L447
(1994).

\bibitem{shif}  M. A. Shifman and A. V. Turbiner, {\em Communications in
Math. Phys.} {\bf 126}, 347 (1989).

\bibitem{ul}  V. V. Ulyanov and O. B. Zaslavskii, {\em Physics Reports} {\bf %
216}, 179 (1992).

\bibitem{mor}  A. Yu. Morozov, A. M. Perelomov, A. A. Rosly, M. A. Shifman
and A. V. Turbiner, {\em Int. J. Mod. Phys.} {\bf A 5}, 803 (1990).
\end{references}

\begin{references}
\bibitem{tag}  Fake bibitem.
\end{references}
\end{document}